\documentclass[review, numbers]{elsarticle} 
\usepackage{fullpage}
\usepackage{amsfonts,amsmath,amssymb,amsthm}
\usepackage{graphicx,psfrag,epsf,comment}
\usepackage{enumerate}
\usepackage{url}
\usepackage{algorithm}
\usepackage{algpseudocode}
\usepackage{helvet}
\usepackage[colorlinks,citecolor=blue]{hyperref}
\usepackage{paralist}
\usepackage{lineno}
\usepackage{chngcntr}
\usepackage{thmtools, thm-restate}

\theoremstyle{plain}

\newtheorem{theorem}{Theorem}
\newtheorem*{theorem*}{Theorem}

\begin{document}

\begin{frontmatter}

 \title{Synergistic Graph Fusion via Encoder Embedding}
\author[1]{Cencheng Shen\corref{cor1}}
\cortext[cor1]{Corresponding author}
\ead{shenc@udel.edu}
\affiliation[1]{organization={University of Delaware},
postcode={19716},
city={Newark},
country={United States}}

\author[2]{Carey Priebe}
\ead{cep@jhu.edu}
\affiliation[2]{organization={Johns Hopkins University},
postcode={21218},
city={Baltimore},
country={United States}}

\author[3]{Jonathan Larson}
\ead{jolarso@microsoft.com}
\affiliation[3]{organization={Microsoft Research},
postcode={98052},
city={Redmond},
country={United States}}

\author[3]{Ha Trinh}
\ead{trinhha@microsoft.com}

\begin{abstract}
In this paper, we introduce a method called graph fusion embedding, designed for multi-graph embedding with shared vertex sets. Under the framework of supervised learning, our method exhibits a remarkable and highly desirable synergistic effect: for sufficiently large vertex size, the accuracy of vertex classification consistently benefits from the incorporation of additional graphs. We establish the mathematical foundation for the method, including the asymptotic convergence of the embedding, a sufficient condition for asymptotic optimal classification, and the proof of the synergistic effect for vertex classification. Our comprehensive simulations and real data experiments provide compelling evidence supporting the effectiveness of our proposed method, showcasing the pronounced synergistic effect for multiple graphs from disparate sources.
\end{abstract}


\end{frontmatter}

\section{Introduction}
Graphs have become increasingly prevalent in various real-world scenarios, including social networks, communication networks, webpage hyperlinks, and biological systems \cite{GirvanNewman2002, newman2003structure, barabasi2004network, boccaletti2006complex, VarchneyEtAl2011, ugander2011anatomy}. In graph data, we have a set of vertices denoted as $\{v_i, i=1,\ldots,n\}$, and edges represented as $\{e_j, j =1,\ldots, s\}$ connecting these vertices. This information can be captured by an $n \times n$ adjacency matrix $\mathbf{A}$, where $\mathbf{A}(i,j)=1$ indicates the existence of an edge between vertices $i$ and $j$, and $\mathbf{A}(i,j)=0$ indicates the absence of an edge. 

Traditionally, the adjacency matrix is binary and primarily used for representing unweighted graphs. Its capacity extends to handling weighted graphs, where the elements of $\mathbf{A}$ are assigned the corresponding edge weights. Moreover, graph representation can be used for any data via proper transformations. For instance, when vertex attributes are available, they can be transformed into a pairwise distance or kernel matrix, thus creating a general graph structure. This adaptability enables the incorporation of additional information from vertex attributes, making graph representation a powerful tool for data analysis that goes beyond traditional binary or weighted graphs.

Graph embedding is a fundamental and versatile approach for analyzing and exploring graph data, encompassing a wide range of techniques such as spectral embedding \cite{RoheEtAl2011,SussmanEtAl2012}, graph convolutional neural networks \cite{kipf2017semi, Wu2019ACS}, node2vec \cite{grover2016node2vec, node2vec2021}, among others. By projecting the vertices into a low-dimensional space while preserving the structural information of the graph, graph embedding yields vertex representations in Euclidean space that facilitate various downstream inference tasks, such as community detection \cite{KarrerNewman2011, ZhaoLevinaZhu2012}, vertex classification \cite{perozzi2014deepwalk, kipf2017semi}, outlier detection \cite{Ranshous2015, akoglu2015graph}, etc.

This manuscript addresses a crucial and increasingly prevalent scenario in data analysis where we encounter multiple graphs $\{\mathbf{A}_1, \mathbf{A}_2, \ldots, \mathbf{A}_{M}\}$ sharing a common vertex set \cite{Park2013, LyzinskiFishkindPriebe2014, ManifoldPRL, arroyo2021inference}. This situation is becoming more frequent as data collection accelerates, and it arises in diverse fields such as social network analysis, neuroscience, and linguistics. For instance, one may want to analyze different social network connections for the same group of individuals, different measurements on the same brain regions, hyperlinks in articles written in different languages, or citation networks enriched with additional attributes. In such settings, it is essential to develop a method that can effectively integrate data from multiple sources. The ideal method should not only exhibit improved performance as more signals from different graphs are incorporated but also be robust against noisy or irrelevant data sources. By leveraging information from multiple graphs, we aim to achieve a richer and more comprehensive representation of the underlying structure and relationships within the data. This can lead to better insights and enhanced accuracy in various downstream tasks, such as vertex classification and link prediction.

Existing works on multiple-graph settings are relatively limited. Most of these investigations, particularly those with theoretical foundations, are based on spectral embedding. For instance, the Omnibus method \cite{PriebeMarchette2012, lyzinski2017fast} constructs an $Mn \times Mn$ matrix by placing the input adjacency on the diagonals and computing the average of two adjacency matrices on each off-diagonal $n \times n$ sub-matrix. This matrix is then projected into an $n \times d$ matrix using spectral embedding. The more recent multiple adjacency spectral embedding (MASE) \cite{arroyo2021inference} method conducts individual spectral embedding for each adjacency matrix, concatenating the resulting embeddings followed by single value decomposition to obtain an $n \times d$ embedding. Another method, unfolded spectral embedding (USE) \cite{Patrick2022}, concatenates the adjacency matrices into an $n \times Mn$ matrix, then applies spectral embedding to obtain an $n \times Md$ embedding. The main difference among these methods lies in how they concatenate the multiple graphs in a proper manner. As they all utilize spectral embedding, specifically singular value decomposition, they require the selection of the dimension $d$. While they have demonstrated certain theoretical properties, the theoretical investigations do not focus on the subsequent downstream task and how well it benefits from collecting more graphs.

To that end, in this paper we present a supervised multi-graph embedding approach called "graph fusion embedding," which builds upon the recently proposed graph encoder embedding \cite{GEE1,GEEDynamics,GEEDistance}. The proposed method is designed to accommodate multiple-graph of distinct structures, including binary, weighted, or general graphs via distance or kernel transformation of vertex attributes. A notable feature of the proposed method is the "synergistic effect" in supervised learning. This effect guarantees that, for sufficiently large vertex sizes, the accuracy of vertex classification consistently benefits from the inclusion of additional graphs and never degrades. We prove this synergistic effect under both the stochastic block model for binary graphs and a more general graph model. Additionally, we explore the asymptotic behavior of graph fusion embedding and establish a sufficient condition for it to achieve asymptotically perfect vertex classification.

To validate the superiority of our proposed method, we conduct extensive simulations and evaluations on real graph data. By applying the graph fusion embedding to various types of graphs, we showcase its effectiveness and versatility in different settings. Our method consistently outperforms existing techniques and demonstrates the synergistic effect, making it a robust and powerful tool for multi-graph analysis. The MATLAB code for the method and simulations are made available on Github\footnote{\url{https://github.com/cshen6/GraphEmd}}. The appendix includes additional simulations validating the proposed methods, all benchmark results using competitor methods, as well as all theorem proofs. 

\section{Method}
\label{sec:method}

\subsection{Main Algorithm}
The graph fusion embedding takes as input $M$ graphs with a common set of $n$ vertices and a label vector representing $K$ communities. For vertices with unknown labels, the corresponding entry in the label vector is set to $0$, and such labels are not used in the embedding process.

\begin{itemize}
\item \textbf{Input}: The graphs $\{\mathbf{A}_{m} \in \mathbb{R}^{n \times n}, m=1,\ldots,M\}$ and a label vector $\mathbf{Y} \in \{0,1,\ldots,K\}^{n}$.
\item \textbf{Step 1}: For each $k=1,\ldots,K$, compute the number of observations per-class 
\begin{align*}
n_k = \sum_{i=1}^{n} I(\mathbf{Y}_i=k).
\end{align*}
\item \textbf{Step 2}: Compute the one-hot encoding matrix $\mathbf{W} \in \mathbb{R}^{n \times K}$ based on $\mathbf{Y}$, then normalize by the number of observations per-class. Specifically, for each vertex $i=1,\ldots,n$, we set
\begin{align*}
\mathbf{W}(i, k) = 1 / n_k
\end{align*} 
if and only if $\mathbf{Y}_i=k$, and $0$ otherwise. Note that vertices with unknown labels are unused and effectively zero, i.e., $\mathbf{W}(i, :)$ is a zero vector if $\mathbf{Y}_i=0$. 
\item \textbf{Step 3}: For each graph $m=1,\ldots, M$, calculate the graph encoder embedding:
\begin{align*}
\mathbf{Z}_{m}=\mathbf{A}_{m} \times \mathbf{W}.
\end{align*}
\item \textbf{Step 4}: Let $\mathbf{Z}_{m}(i, \cdot)$ denote each row of $\mathbf{Z}_{m}$, and $\|\cdot\|$ be the Euclidean norm. For each $i$ and each $m$ where $\|\mathbf{Z}_{m}(i, \cdot)\| >0$, normalize the non-zero row by the norm
\begin{align*}
\mathbf{\tilde{Z}}_{m}(i, \cdot)=\frac{\mathbf{Z}_{m}(i, \cdot)}{\|\mathbf{Z}_{m}(i, \cdot)\|},
\end{align*}
then construct the row-concatenated embedding $\mathbf{\tilde{Z}}$ by setting
\begin{align*}
\mathbf{\tilde{Z}}(i, \cdot)=[\mathbf{\tilde{Z}}_{1}(i, \cdot), \mathbf{\tilde{Z}}_{2}(i, \cdot), \ldots, \mathbf{\tilde{Z}}_{M}(i, \cdot)] \in \mathbf{R}^{MK}
\end{align*}
for each vertex $i$.
\item \textbf{Output}: The graph fusion embedding $\mathbf{\tilde{Z}} \in \mathbb{R}^{n \times MK}$.
\end{itemize}

\subsection{Subsequent Classification}
In the supervised setting, the label vector is always available. During steps 1 to 3 of the graph fusion embedding process, the labels of training vertices are used to construct the $W$ matrix, while the labels of testing vertices are set to zero and remain unused throughout the embedding process. It is important to note that both the training and testing vertices are present in the graph, and as a result, the resulting embedding $\mathbf{\tilde{Z}}$ provides representations for all vertices in the graph. Let $\mathbf{Y}_i$ be a random variable taking values in $1,\ldots,K$ that represents the true label for any testing vertex $i$, and let $g(\cdot)$ denotes a classifier function. The true classification error can be denoted as
\begin{align*}
L=\text{Prob}(g(\mathbf{\tilde{Z}}(i,:)) \neq \mathbf{Y}_i).
\end{align*}

Throughout the main paper, we employed the 5-nearest-neighbor classifier as our classifier of choice. The synergistic effect is also observed when using other standard classifiers, such as discriminant analysis or a simple two-layer neural network. These alternatives classifiers can be readily applied, yielding similar numerical outcomes, and are presented in the appendix. Furthermore, the main algorithm can be adapted to an unsupervised version in an iterative manner, similar to the single graph version in \cite{GEEClustering}, which is also briefly discussed in the appendix.

\subsection{Computational Complexity}

The computation complexity of the graph fusion embedding is $O(nMk+\sum_{m=1}^{M}s_{m})$, where $s_{m}$ is the number of entries in each graph. Step 1 and 2 are shared across all graphs, and step 3 is carried out separately for each graph, allowing for parallelization and further reducing the running time to $O(nk +\max_{m=1,\ldots,T}\{s_{m}\})$. Note that if $M=1$, the fusion embedding is equivalent to the normalized graph encoder embedding for a single graph. Additionally, for sparse graphs, step 3 can be implemented in linear time using two simple edgelist operations. 

\begin{figure}[htbp]
	\centering
	\includegraphics[width=1.0\textwidth,trim={0cm 0.5cm 2cm 0.5cm},clip]{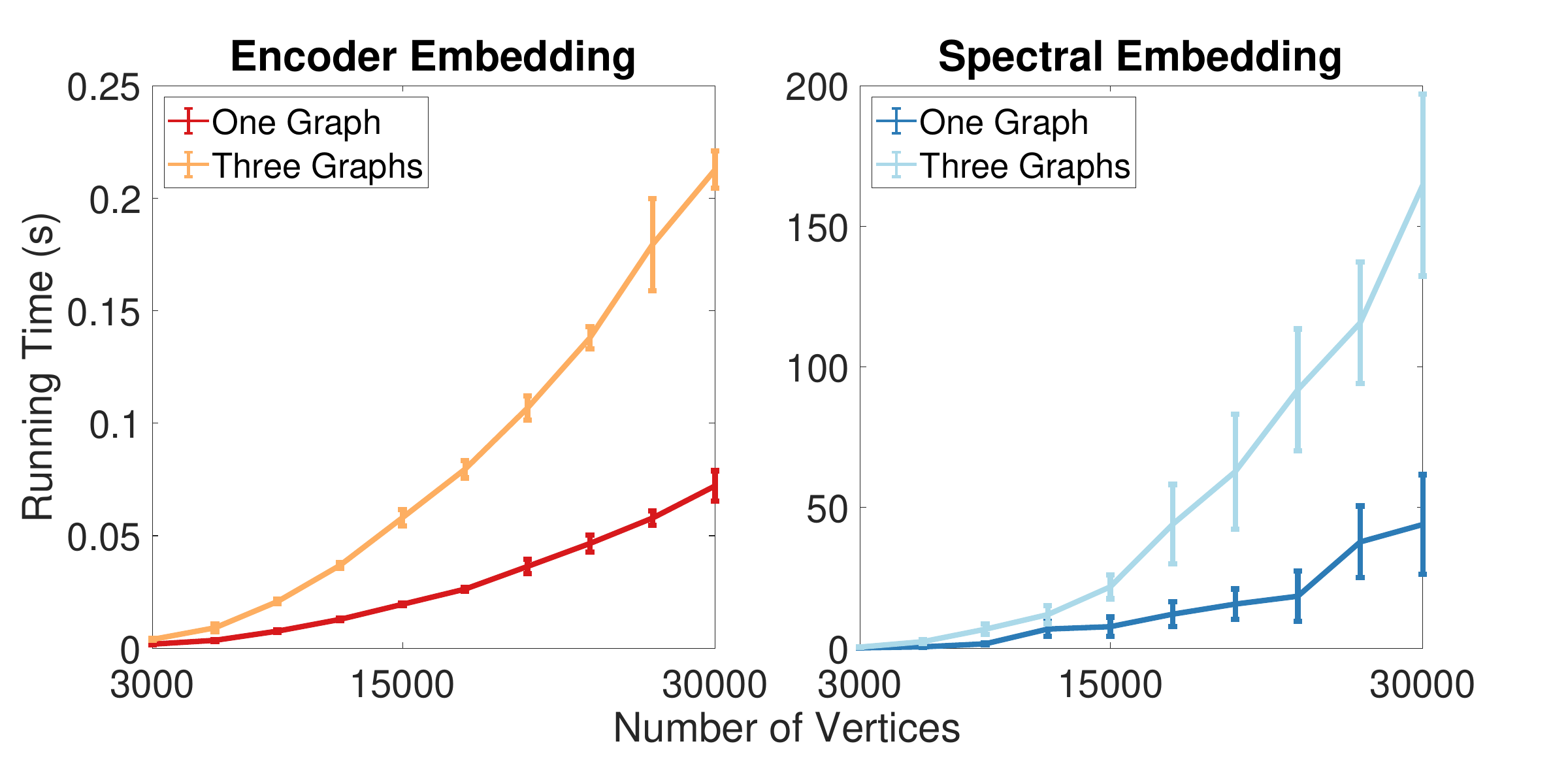}
	\caption{This figure illustrates the runtime performance of the graph fusion embedding (left panel) and the unfolded spectral embedding (right panel). The graphs are generated by simulation model 2 in Section~\ref{sec:sim}, and we carry out $20$ Monte-Carlo replicates at each sample size $n$. Each line represents the average running time, with error bars indicating the standard deviation.}
	\label{figS10}
\end{figure}

Figure~\ref{figS10} provides a comparison of the running time between the graph fusion embedding and the unfolded spectral embedding. The algorithm demonstrates linear scalability with respect to the number of graphs, with the three-graph fusion embedding taking approximately three times longer than the single-graph embedding. Additionally, the graph fusion embedding is significantly more efficient than the spectral embedding, requiring only $0.2$ seconds at $n=30000$, whereas the spectral embedding takes over $4$ minutes. In general, the encoder embedding outperforms any existing graph embedding approach and can process millions of edges in just a few seconds. 

\section{Theory on Binary Graph}

\subsection{The Stochastic Block Model}
\label{sec1}
The binary graph serves as the foundation of most graph methods and is commonly used in various applications. To demonstrate the functionality of graph fusion embedding on binary graphs, we focus on the stochastic block model (SBM) for both theoretical analysis and simulations.

Under SBM, each vertex $i$ in a graph $\mathbf{A}_{m}$ is assigned a label $\mathbf{Y}_i \in \{1,\ldots, K\}$. The probability of an edge between a vertex from class $k$ and a vertex from class $l$ is determined by a block probability matrix $\mathbf{B}_{m}=[\mathbf{B}_{m}(k,l)] \in [0,1]^{K \times K}$, and for any $i \neq j$ it holds that
\begin{align*}
\mathbf{A}_{m}(i,j) &\sim \operatorname{Bernoulli}(\mathbf{B}_{m}(\mathbf{Y}_i, \mathbf{Y}_j)).
\end{align*}
It is worth mentioning that this formulation generates a directed graph, but it can be easily adapted for an undirected graph by setting $\mathbf{A}_{m}(j,i) = \mathbf{A}_{m}(i,j)$ for all $i<j$. The choice of directed or undirected graph does not affect the method or theory, and in this case, we opt for the directed case. Additionally, the graph may or may not have self-loops, i.e., $\mathbf{A}_{m}(i,i)$ could be $0$ or $1$, and the method and the theorems still hold.

The degree-corrected stochastic block model (DC-SBM) is an extension of the SBM that accounts for the sparsity observed in real graphs \cite{ZhaoLevinaZhu2012}. In addition to the parameters defined in the SBM, each vertex $i$ is associated with a degree $\theta_i \stackrel{i.i.d.}{\sim} F_\theta$. When conditioned on the degrees, each edge from vertex $i$ to another vertex $j$ is independently generated as follows:
\begin{align*}
\mathbf{A}_{m}(i,j) \sim \operatorname{Bernoulli}(\theta_i \theta_j \mathbf{B}(\mathbf{Y}_i, \mathbf{Y}_j)).
\end{align*}
As standard SBM is a special case where all degrees are $1$, it suffices to consider the DC-SBM graphs for theory purpose. 

Finally, the class labels $\mathbf{Y}_i$ are assumed to follow a categorical distribution with prior probability $\{\pi_k \in (0,1) \mbox{ with } \sum_{k=1}^{K} \pi_k=1\}$. Under this assumption, as the number of vertices $n$ increases to infinity, the number of vertices within each class $n_k$ also converges to infinity. If not, $\pi_k$ asymptotically approaches $0$, rendering class $k$ trivial in the model. This assumption implies that $n_k = O(n)$ for all $k$.

Note that we do not assume any specific dependence structure between graphs. For instance, $\mathbf{A}_{1}(i,j)$ and $\mathbf{A}_{2}(i,j)$ could be independently distributed, identical, similar up to some noise or transformation, or correlated in any other way, and all theorems would still hold true.

\subsection{Convergence of the Fusion Embedding}
Given $M$ graphs from DC-SBM with a common vertex set, we define $\mathbf{\tilde{B}}_{m}$ as the row-normalized block probability matrix of $\mathbf{B}_{m}$, such that for each row $k$, 
\begin{align*}
\mathbf{\tilde{B}}_{m}(k,:) = \frac{\mathbf{B}_{m}(k,:)}{\|\mathbf{B}_{m}(k,:)\|}.
\end{align*}
Then we form the row concatenation of all $\mathbf{\tilde{B}}_{m}$ as:
\begin{align*}
\mathbf{\tilde{B}} = [\mathbf{\tilde{B}}_{1}, \mathbf{\tilde{B}}_{2},\ldots, \mathbf{\tilde{B}}_{M}] \in \mathbb{R}^{K \times MK}.
\end{align*}
The first theorem demonstrates that given DC-SBM graphs, the graph fusion embedding converges to the normalized and concatenated block probability matrix.
\begin{theorem}
\label{thm1}
Suppose that $\{\mathbf{A}_{m}, m=1,\ldots,M\}$ follows the DC-SBM model in Section~\ref{sec1}. Let $n$ be the number of vertices with known labels. Then, for any vertex $i$ belonging to class $\mathbf{Y}_i$, its graph fusion embedding satisfies:
\begin{align*}
\|\mathbf{\tilde{Z}}(i, \cdot) - \mathbf{\tilde{B}}(\mathbf{Y}_i,:) \| = O(\frac{1}{\sqrt{n}}) \stackrel{n \rightarrow \infty}{\rightarrow} 0.
\end{align*}
\end{theorem}

\subsection{Asymptotic Perfect Separation}
Under DC-SBM, the normalized and concatenated block probability matrix $\mathbf{\tilde{B}}$ fully determines the embedding. As a result, there exists a simple condition to ensure the embedding is asymptotically perfect for vertex classification. In the context of vertex classification, both the training and testing vertices follow the same asymptotic convergence in Theorem~\ref{thm1}, with the only difference being that the training labels are known while the testing labels are not observed. We denote the finite-sample nearest-neighbor classification error using $M$ graphs and $n$ training vertices as $L_{n}(\mathbf{A}_{1},\ldots,\mathbf{A}_{M})$.
\begin{theorem}
\label{thm2}
Suppose that $\{\mathbf{A}_{m}, m=1,\ldots,M\}$ follows the DC-SBM model in Section~\ref{sec1}. As the number of training vertices increases to infinity, the graph fusion embedding achieves asymptotically perfect vertex classification, i.e., 
\begin{align*}
\lim_{n\rightarrow \infty }L_{n}(\mathbf{A}_{1},\ldots,\mathbf{A}_{M}) = 0,
\end{align*}
if and only if there are no repeating rows in $\mathbf{\tilde{B}}$.
\end{theorem}
In essence, the graph fusion embedding can achieve perfect classification when there is no overlap between the rows of $\mathbf{\tilde{B}}$, meaning that each row represents a unique class. Conversely, if any two rows of $\mathbf{\tilde{B}}$ coincide, then the embedding vectors for vertices in those two classes will also coincide, making them indistinguishable.

While the condition may appear restrictive, it is equivalent to stating that every class of vertices must be perfectly separable from any other class, a scenario that rarely occurs in practice. Indeed, perfect classification serves as a preliminary step for the subsequent theorem on the synergistic effect, which always holds for our fusion embedding and does not require additional conditions.


\subsection{The Synergistic Effect}
\begin{theorem}
\label{thm3}
Suppose that $\{\mathbf{A}_{m}, m=1,\ldots,M\}$ follows the DC-SBM model in Section~\ref{sec1}. For any subset of graphs $M_1 \leq M$, the classification error using the fusion embedding satisfies:
\begin{align*}
\lim_{n\rightarrow \infty }L_{n}(\mathbf{A}_{1},\ldots,\mathbf{A}_{M}) - L_{n}(\mathbf{A}_{1},\ldots,\mathbf{A}_{M_1}) \leq 0
\end{align*}
for sufficiently large sample size $n$.
\end{theorem}

Therefore, as more graphs are collected and included, vertex classification via the graph fusion embedding always achieves equal or better performance. If an included graph contains pure noise, the classification error does not deteriorate; however, any graph that contains additional signal to better separate classes can lead to a reduction in classification error.
It is important to note that the theorems are based on asymptotic analysis, and the finite-sample error may still exhibit minimal deterioration, especially for small vertex size.

\section{Extension to General Graphs}

\subsection{A General Graph Model}
\label{sec2}
To understand how the graph fusion embedding can be applied to weighted graphs or distance transformation of vertex attributes, we consider the following general graph model. For any fixed vertex $i$, we assume that 
\begin{align*}
\mathbf{A}_{m}(i,1), \mathbf{A}_{m}(i,2), \cdots &\stackrel{i.i.d.}{\sim} F_{m}(i),
\end{align*}
and the densities $\{F_{m}(i), m=1,\ldots,M \mbox{ and } i=1,\ldots,n\}$ have finite moments. Namely, each row of the adjacency matrix is independently and identically distributed when conditioning on vertex $i$. Note that when we condition on the vertex $i$, it means that all relevant information associated with that vertex are known, including its degree, class label, or Euclidean position, depending on the underlying distribution of the entries in the adjacency matrix.

This model includes a wide range of graph representations, including the aforementioned degree-corrected stochastic block model (DC-SBM), the random dot product graph model \cite{YoungScheinerman2007}, as well as a pairwise distance or kernel matrix from Euclidean space. To illustrate, consider the scenario where we have vertex attributes $X_i$ for each vertex $i=1,\ldots, n$, and we construct a Euclidean distance matrix $\mathbf{A}$ where $\mathbf{A}(i,j)=\|X_i-X_j\|$. Under this transformation, when we condition on the attributes $X_i$, the entries $\mathbf{A}(i,j)$ and $\mathbf{A}(i,s)$ are independently and identically distributed for any $j\neq s$. 

We reiterate that the general model does not assume any dependence structure between graphs. For instance, $\mathbf{A}_{1}(i,j)$ and $\mathbf{A}_{2}(i,j)$ could be correlated in any manner. The model assumption is solely on the edge generation within each individual graph.

\subsection{Convergence under General Graph}
Given the above model, we define the following conditional expectation for the $m$th graph:
\begin{align*}
E(\mathbf{A}_{m}(i,j) | i)=[E(\mathbf{A}_{m}(i,j) | i, \mathbf{Y}_j=1), \cdots, E(\mathbf{A}_{m}(i,j) | i, \mathbf{Y}_j=K)],
\end{align*}
which is a K-dimensional vector. We then define the normalized vector as follows:
\begin{align*}
\mu_{m}(i) = \frac{E(\mathbf{A}_{m}(i,j) | i)}{\|E(\mathbf{A}_{m}(i,j) | i)\|} \in \mathbb{R}^{K}.
\end{align*}
Finally, we form the concatenated mean matrix as follows:
\begin{align*}
\mu(i)=[\mu_{1}(i), \mu_{2}(i), \ldots, \mu_{M}(i)] \in \mathbb{R}^{MK},
\end{align*}
which represents the concatenation of all the normalized conditional expectation vectors for vertex $i$ across all $M$ graphs.

\begin{theorem}
\label{thm4}
Suppose that $\{\mathbf{A}_{m}, m=1,\ldots,M\}$ follows the general graph model in Section~\ref{sec2}. Let $n$ be the number of vertices with known labels. Then, for any vertex $i$ belonging to class $\mathbf{Y}_i$, its graph fusion embedding satisfies:
\begin{align*}
\|\mathbf{\tilde{Z}}(i, :) - \mu(i) \| = O(\frac{1}{\sqrt{n}}) \stackrel{n \rightarrow \infty}{\rightarrow} 0.
\end{align*}
\end{theorem}

Thus, in the context of a general graph model, the fusion embedding provides $n$ point masses represented by the concatenated conditional means. The location of each point mass is determined by the specific vertex $i$ under consideration, reflecting the characteristics of the underlying graph model. For example, under the DC-SBM graph, $\mu(i)=\mathbf{\tilde{B}}(\mathbf{Y}_i,:)$, implying that all vertices belonging to the same class share a common point. This effectively reduces $n$ embedding position to $K$ points, where each point corresponds to a distinct class. 

\subsection{Vertex Classification under General Graph}
Let $\{h_{k}(x), k=1,\ldots,K\}$ denotes the density family where each $h_{k}(x)$ represents all possible $\mu(i)$ from class $k$. Generally, $h_{k}(x)$ is continuous, but in specific cases, such as discrete graph models, the density can be discrete, e.g., under DC-SBM, $h_{k}(\mathbf{\tilde{B}}(k,:))=1$ and $0$ otherwise. Using the density characterization, the next theorem expands the results of Theorem~\ref{thm2} and Theorem~\ref{thm3} to the general graph model. It establishes a general condition for achieving asymptotically optimal classification through the fusion embedding and demonstrates the presence of a synergistic effect in general scenarios. 

\begin{theorem}
\label{thm5}
Suppose that $\{\mathbf{A}_{m}, m=1,\ldots,M\}$ follows the general graph model in Section~\ref{sec2}. As the number of training vertices increases to infinity, the graph fusion embedding achieves asymptotically perfect vertex classification, i.e., 
\begin{align*}
\lim_{n\rightarrow \infty }L_{n}(\mathbf{A}_{1},\ldots,\mathbf{A}_{M}) = 0
\end{align*}
if and only if 
\begin{align*}
\sum\limits_{1 \leq k < l \leq K}h_{k}(x)h_{l}(x) = 0
\end{align*}
for any $x \in \mathbb{R}^{MK}$.

Additionally, for any subset of graphs $M_1 \leq M$, the classification error using the fusion embedding satisfies:
\begin{align*}
\lim_{n\rightarrow \infty }L_{n}(\mathbf{A}_{1},\ldots,\mathbf{A}_{M_1},\ldots,\mathbf{A}_{M}) - L_{n}(\mathbf{A}_{1},\ldots,\mathbf{A}_{M}) \leq 0
\end{align*}
for sufficiently large sample size $n$.
\end{theorem}

Once more, the condition for asymptotically perfect classification implies that each class must be perfectly separable, as indicated by the class-wise density of the expectations. Although such perfect separability is not commonly observed, it serves as an important special case in understanding the synergistic effect.

Note that based on the convergence rate for the embedding ($O(\frac{1}{\sqrt{n}})$ as shown in Theorem~\ref{thm4}), it is possible to estimate the required sample size to observe the classification improvement. However, the precise value of $n$ is contingent not only on the convergence rate but also on factors like the magnitude of separation between classes and the standard deviation of the underlying graph distribution.

As an illustration, let us make some simplifying assumptions. Suppose each class-conditional mean (i.e., $\mu(i)$) is at least $c_1$ apart from others, and the standard deviation for the graph variable (i.e., $F_{m}(i)$) is $c_2$. In this case, we can employ a normal approximation with three standard deviations, and deduce that the required sample size for classification error convergence must satisfy $6 \frac{c_2}{\sqrt{n}} \geq c_1$.

For example, considering $c_1=0.1$ and $c_2 = 0.5$, solving the inequality yields $n \geq 900$, which is similar to the setting for the rightmost figure in Figure~\ref{fig1} with one graph. As another example, if we assume $c_1=0.02$ (indicating very small differences in class-conditional expectations) and $c_2 = 0.4$, then $n$ would need to be approximately $15000$ to achieve perfect classification.

\section{Simulations}
\label{sec:sim}

In this section, we present three simulations to illustrate the synergistic effect of the graph fusion embedding. We consistently use the 5-nearest-neighbor classifier in all simulations and real data experiments. While the main paper primarily focuses on the graph fusion embedding and the synergistic effect in classification, additional simulation results are provided in the appendix. These include experiments with heterogeneous graphs, alternative classifiers, the unsupervised case, and comparisons with benchmark methods (Omnibus, MASE, and USE). 

\subsection*{Simulation 1}
\begin{figure}[htbp]
	\centering
 \includegraphics[width=1.0\textwidth,trim={0cm 0cm 0cm 0cm},clip]{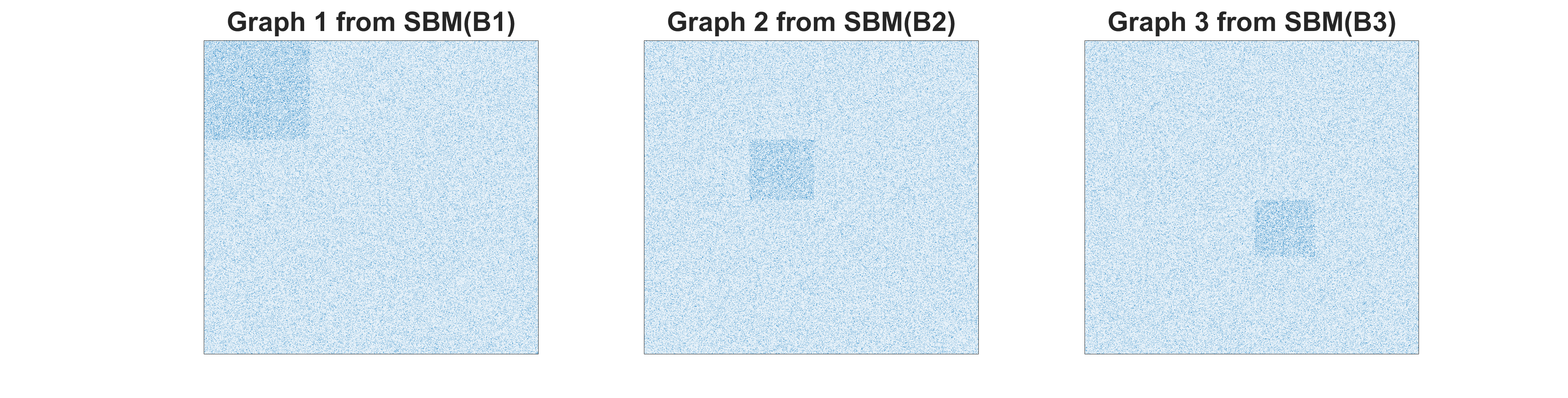}
	\caption{Visualize three SBM graphs at $n=1000$. }
	\label{fig0}
\end{figure}

We consider a scenario with $K=4$ classes, where the prior probabilities for each vertex to belong to class $[1,2,3,4]$ are set to $[0, 0.3, 0.2, 0.2, 0.3]$. Three SBM graphs are generated with block probability matrices $B_{m} \in \mathbb{R}^{K \times K}$. For each $m$, the probabilities are $B_{m}(k,l)=0.1$ for all $k, l$, except $B_{m}(m,m)=0.2$ for $m=1,2,3$. This means that the $m$th class has a stronger connection in graph $m$. A visualization of this setup is provided in Figure~\ref{fig0}. While no single graph is asymptotically perfect for classification, the concatenated block probability matrix
\begin{align*}
\mathbf{B}=
  \left[\begin{matrix}
    0.2 & \cdots & 0.1 & 0.1 & \cdots & 0.1 & 0.1 & 0.1 & \cdots\\
    0.1 & \cdots & 0.1 & 0.2 & \cdots & 0.1 & 0.1 & 0.1 & \cdots\\
    0.1 & \cdots & 0.1 & 0.1 & \cdots & 0.1 & 0.1 & 0.2 & \cdots \\
    0.1 & \cdots & 0.1 & 0.1 & \cdots & 0.1 & 0.1 & 0.1 & \cdots 
  \end{matrix}\right]
\end{align*}
satisfies the condition in Theorem~\ref{thm2}. Therefore, each graph provides additional information to aid in separating the communities, and using all three graphs leads to asymptotic perfect classification.

\subsection*{Simulation 2}
Simulations 2 uses the same settings as simulation 1, except that the three graphs are generated using DC-SBM. The parameters are the same as above, and the degrees are drawn independently from a uniform distribution in the interval $[0.1, 0.5]$.

\subsection*{Simulation 3}
This simulation involves starting with one signal graph and progressively adding independent graphs. The number of classes remains at four with the same prior probability. We first generate $\mathbf{A}_{1}$ using SBM with the block probability matrix $\mathbf{B}_{1}$, which has $0.1$ in off-diagonal entries and $0.2$ in diagonal entries. Next, we generate five independent SBM graphs $\mathbf{A}_{m}$ for $m=2,\ldots, 6$ with the block probability matrix $\mathbf{B}_{m}$ being $0.1$ in all entries. The concatenated block probability matrix $\mathbf{B}$ always satisfies Theorem~\ref{thm2}, even in the presence of independent graphs, i.e., 
\begin{align*}
\mathbf{B}=
  \left[\begin{matrix}
    0.2 & 0.1 & 0.1 & 0.1 & \cdots\\
    0.1 & 0.2 & 0.1 & 0.1 & \cdots \\
    0.1 & 0.1 & 0.2 & 0.1 & \cdots \\
    0.1 & 0.1 & 0.1 & 0.2 & \cdots 
  \end{matrix}\right]
\end{align*}
Theorems~\ref{thm2} and~\ref{thm3} predict that the classification error should stay almost the same with the addition of noisy graphs.

\subsection*{Results}
\begin{figure}[htbp]
	\centering
	\includegraphics[width=1.0\textwidth,trim={0cm 0.5cm 2cm 0.5cm},clip]{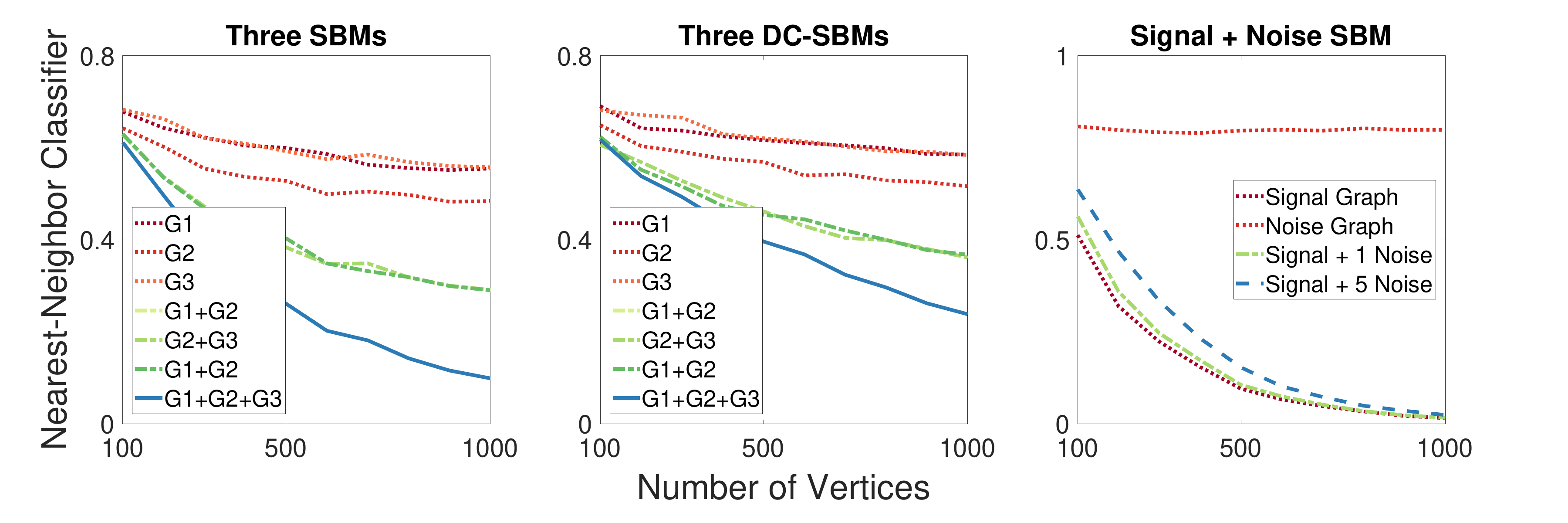}
	\caption{The figure shows the 5-fold classification error of graph fusion embedding, averaged over 20 Monte-Carlo replicates, for the three simulation settings.}
	\label{fig1}
\end{figure}

In both simulation 1 and simulation 2, the classification error improves with the inclusion of more graphs, as illustrated in the left and center panels of Figure~\ref{fig1}. Using two graphs yields a lower error than using only one graph, and using all three graphs yields the lowest error. This is because each new graph contains new signal about the labels, so including a new graph shall always yields better performance.

In simulation 3, only the first graph is related to the label, while all additional graphs are essentially noise. The addition of noise graphs has a minimal effect on the classification error of fusion embedding as the number of vertices increases, as shown in the right panels of Figure~\ref{fig1}. Compared to the benchmark methods' performance in the appendix, the graph fusion embedding stands out as the only method that demonstrates the synergistic effect, remains robust against noise graphs, and consistently performs well across all simulations.

\section{Real Data}

In this section, we showcase the performance and synergistic effect of the graph fusion embedding using a wide range of real-world graphs. The dataset consists of various types of graphs, including binary graphs, weighted graphs, directed and undirected graphs, as well as general graphs such as distance matrices of vertex attributes. The majority of the data were sourced from the network repository \cite{nr}, while a few were collected directly by the authors. All the public data has been properly formatted in MATLAB and is made available on GitHub\footnote{\url{https://github.com/cshen6/GraphEmd}}. The results for Omnibus, MASE, and USE methods are provided in the appendix for comparison.

\subsection{Two-Graph Data}

Table~\ref{table1} presents the 5-fold classification error for five graph datasets, each having two graphs. The results are reported as the mean and standard deviation over $20$ Monte-Carlo replicates. A brief description of each data and pre-processing steps is as follows:
\begin{itemize}
\item The C-elegans neuron data \cite{PavlovicEtAl2014} consists of two binary graphs with $253$ vertices, over $1000$ edges, and a label vector with three classes. 
\item The cora dataset \cite{mccallum2000automating} is a citation network with $2708$ vertices and $5429$ edges and $7$ classes, and each vertex has a $1433$-dimensional 0/1 vertex attribute, indicating the absence or presence of corresponding words from the dictionary. $\mathbf{A}_1$ is a binary matrix indicating the existence of edges, while $\mathbf{A}_2=\mathbf{1}-\mathbf{D}$ where $\mathbf{D}$ is the pairwise cosine distance matrix of the vertex attributes and $\mathbf{1}$ is the matrix of ones.
\item The citeseer data \cite{giles1998citeseer} is another citation network with $3312$ vertices and $4715$ edges and $6$ classes, with a vertex attribute of $3703$ dimensions. We apply the same processing as in the cora dataset to yield a binary graph $\mathbf{A}_1$ and a general graph $\mathbf{A}_2=\mathbf{1}-\mathbf{D}$ from the cosine distance of the attributes.
\item The IMDB \cite{yanardag2015deep} data consists of two slighly un-matched graphs, one with binary labels and $19773$ vertices, and another with three-class labels and $19502$ vertices. We remove the $271$ unmatched vertices to ensure a common vertex set, and use the $3$-class label as the label vector.
\item COIL-RAG has $11757$ vertices with $23588$ edges, $100$ classes, and a vertex attribute of $64$ dimensions. $\mathbf{A}_1$ is a binary matrix indicating the existence of edges, while $\mathbf{A}_2=\max\{\mathbf{D}\}-\mathbf{D}$ where $\mathbf{D}$ is the pairwise Euclidean distance matrix of the vertex attributes. 
\end{itemize}
In the cora and citeseer datasets, we employed the cosine distance as it is a widely-used and effective distance measure in word-frequency analysis \cite{BleiNgJordan2003}. Additionally, we transformed the distance matrix by subtracting the distance entries from the maximum value in the distance matrix, denoted as $\max\{\mathbf{D}\}$ (which equals $1$ for cosine distance). This transformation ensures that the resulting matrix is a valid kernel or similarity matrix \cite{DcorKernel}.

The results presented in Table~\ref{table1} (and compared with appendix Table~\ref{tableA2}) clearly demonstrates that the graph fusion embedding outperforms all other methods by a significant margin across all real datasets, regardless of whether it involves single graph classification, multiple binary graph classification, or multiple general graphs. The only exception is the C-elegans data, where the fusion embedding exhibits slightly lower performance, likely due to the small size of the vertex set. Furthermore, the graph fusion embedding consistently exhibits improved classification performance as more graphs are included. 


\begin{table}[htbp]
\renewcommand{\arraystretch}{1.3}
\centering
{\begin{tabular}{c||c|c|c}
 \hline
 & Graph 1 & Graph 2 & Graph 1+2  \\
\hline
C-elegans & $42.1\% \pm 2.3\%$ & $44.5\% \pm 3.0\%$   & $\mathbf{38.3\%} \pm 2.4\%$  \\
Cora & $19.2\% \pm 0.5\%$ & $29.7\% \pm 0.5\%$     &  $\mathbf{15.8\%} \pm 0.5\%$  \\
Citeseer  & $32.3\% \pm 1.6\%$  & $31.1\% \pm 0.3\%$   &  $\mathbf{26.1\%} \pm 0.5\%$  \\
COIL-RAG  & $7.44\% \pm 0.4\%$ & $48.4\% \pm 0.03\%$    &  $\mathbf{3.72\%} \pm 0.2\%$ \\
IMDB & $0.9\% \pm 2\%$     &  $0.02\% \pm 0.02\%$  & $\mathbf{0\%} \pm 0\%$  \\
\hline
\end{tabular}
\caption{Evaluate the classification errors using graph fusion embedding. We report the average 5-fold errors and the standard deviation over $20$ Monte-Carlo replicates. }
\label{table1}
}
\end{table}

\subsection{Three-Graph and Three-Label Data}

The letter dataset consists of three binary graphs, named letter-high, letter-med, and letter-low, each of which has its own label vector with a total of $15$ classes. The graphs contain over $10500$ vertices and $20250$, $14092$ and $14426$ edges, respectively. After removing several unmatched vertices, we use all three graphs to predict the class labels, and the results are shown in Table~\ref{table2}. The results show that the fusion embedding consistently exhibits the synergistic effect. Specifically, the classification error for each graph and its corresponding label is better than that of a mismatched pair, for example using Letter-High to predict Label-High results in better error than using Letter-Mid for the same task. Additionally, using all three graphs yields the best classification error in all cases. 

The dataset comprises graph representations of $15$ letters: a, e, f, h, i, k, l, m, n, t, v, w, x, y, z. The labels "high," "med," and "low" denote the letter's shape, with "high" indicating capitalized letters. The drawings are distorted, making letter recognition a non-trivial task. In the graph, vertices denote the endpoints of the letter drawing, while edges represent the lines in the drawing. Our results indicate that capitalized letters are the easiest to recognize, likely due to their distinct shapes compared to smaller letters. Moreover, employing the fusion embedding using all three graphs yields the best performance in identifying the letter (or, more precisely, the endpoint of a letter) each vertex belongs to.

\begin{table}[htbp]
\renewcommand{\arraystretch}{1.3}
\centering
{\begin{tabular}{c||c|c|c|c}
 \hline
 & Letter-High & Letter-Med & Letter-Low & All Graphs  \\
\hline
Label-High & $7.86\% \pm 0.3\%$ & $44.2\% \pm 0.8\%$   & $43.0\% \pm 0.9\%$  & $\mathbf{6.09\%} \pm 0.2\%$  \\
Label-Med & $44.6\% \pm 0.9\%$ & $14.4\% \pm 0.4\%$   & $44.1\% \pm 0.6\%$   &  $\mathbf{9.75\%} \pm 0.3\%$  \\
Label-Low & $44.3\% \pm 0.6\%$  & $43.2\% \pm 0.5\%$  & $13.9\% \pm 0.4\%$  &  $\mathbf{9.50\%} \pm 0.2\%$  \\
\hline
\end{tabular}
\caption{Evaluate the classification errors using fusion embedding for a three-graph three-label data.}
\label{table2}
}
\end{table}

\subsection{Four-Graph Wikipedia Data}

The Wikipedia dataset consists of four disparate graph representations we collected from Wikipedia articles \cite{GCCAJMVA,ManifoldPRL}. The dataset has a total of $n=1382$ Wikipedia articles based on a 2-neighborhood of the English article ``Algebraic Geometry" and the corresponding French articles. Latent Dirichlet Allocation is applied to the word-frequency of each article to calculate the English text feature and French text feature, from which we generate two graphs $\mathbf{A}_1$ and $\mathbf{A}_2$ via the cosine distance transformation. The hyperlink information in the English articles and the corresponding French articles yields $\mathbf{A}_3$ and $\mathbf{A}_4$. All articles are manually labeled into $K=5$ disjoint classes: category, people, locations, date, math.

Table~\ref{table3} displays the classification errors using 5-nearest-neighbor on graph fusion embedding for the Wikipedia dataset. For each individual graph, it is evident that the text graphs provide more information for predicting the class of the article, as expected. On the other hand, the hyperlink graph exhibits significantly higher classification errors, indicating less signals. Moreover, combining the text and hyperlink graphs results in slightly improved performance compared to using individual graph. The most favorable outcome is achieved when all four graphs are utilized, showcasing the synergistic effect.

\begin{table}[htbp]
\renewcommand{\arraystretch}{1.3}
\centering
\scalebox{0.9}{
\begin{tabular}{c||c|c|c|c}
\hline
 & English Text & French Text & English Hyperlinks & French Hyperlinks  \\
\hline
One Graph & $19.3\% \pm 0.5\%$  & $18.5\% \pm 0.5\%$ & $48.6\% \pm 0.9\%$ & $52.9\% \pm 1.2\%$ \\
Two Graphs & $16.2\% \pm 0.5\%$ (1+2)   & $18.4\% \pm 0.6\%$ (2+4) & $19.0\% \pm 0.5\%$ (1+3) & $44.7\% \pm 1.0\%$ (3+4) \\
Three Graphs  & $16.2\% \pm 0.5\%$ (-4)  & $16.1\% \pm 0.5\%$ (-3) & $18.8\% \pm 0.5\%$ (-2) & $18.4\% \pm 0.5\%$ (-1) \\
\hline
All & \multicolumn{4}{c}{$\mathbf{16.0\%} \pm 0.5\%$} \\
\hline
\end{tabular}
}
\caption{Evaluate the classification errors using graph fusion embedding for the Wikipedia data. The first row evaluates the embedding of each individual graph; the second row evaluates the embedding of using two graphs, e.g., 1+2 means English text + French text; the third row evaluates the embedding of using three graphs, e.g., -4 means all graphs excluding French hyperlinks; the last row uses all four graphs. }
\label{table3}
\end{table}

\section{Discussion}

This paper presented the graph fusion embedding, and provided compelling evidence on its theoretical soundness and numerical advantages. The most important feature of this method is the synergistic effect, which results in improved numerical performance as more graphs are included. To some extent, the method can be viewed as a supervised and deterministic version of spectral embedding, but without requiring any additional transformation, and works well for general graphs. In contrast, many competitor methods require extra alignment or additional singular value decomposition. 

This paper suggests several promising directions for future research. Firstly, as the dimension of the fusion embedding always equals $MK$, for an extremely large number of graphs with high $M$, or an extremely large number of classes $M$, exploring dimension reduction techniques and strategies for selecting graphs would be intriguing. Secondly, since the graph encoder embedding can be employed for distance and kernel matrices, exploring connections with other areas such as multi-dimensional scaling \cite{BorgBook, CoxBook}, distance and kernel correlation \cite{SzekelyRizzo2009, GrettonEtAl2005, DcorFast}, etc., could provide valuable insights. Lastly, the application of this method to general graphs suggests its potential to enhance existing data fusion tasks, provided an appropriate metric can be identified. Hence, exploring more applications using this method and possible variants on different data regimes would be an exciting avenue for future research.


\section*{Acknowledgment}
This work was supported in part by 
the Defense Advanced Research Projects Agency under the D3M program administered through contract FA8750-17-2-0112, the National Science Foundation HDR TRIPODS 1934979, the National Science Foundation DMS-2113099, and by funding from Microsoft Research.

\bibliographystyle{abbrvnat} 
\bibliography{shen,general}


\clearpage
\appendix
\setcounter{figure}{0}
\setcounter{table}{0}
\setcounter{theorem}{0}
\renewcommand{\thealgorithm}{C\arabic{algorithm}}
\renewcommand{\thefigure}{B\arabic{figure}}
\renewcommand{\thesubsection}{\thesection.\arabic{subsection}}
\renewcommand{\thesubsubsection}{\thesubsection.\arabic{subsubsection}}
\pagenumbering{arabic}
\renewcommand{\thepage}{\arabic{page}}

\begin{center}
{\large\bf APPENDIX}
\end{center}

\section{Theorem Proofs}
\begin{theorem}
Suppose that $\{\mathbf{A}_{m}, m=1,\ldots,M\}$ follows the DC-SBM model in Section~\ref{sec1}. Let $n$ be the number of vertices with known labels. Then, for any vertex $i$ belonging to class $\mathbf{Y}_i$, its graph fusion embedding satisfies:
\begin{align*}
\|\mathbf{\tilde{Z}}(i, \cdot) - \mathbf{\tilde{B}}(\mathbf{Y}_i,:) \| = O(\frac{1}{\sqrt{n}}) \stackrel{n \rightarrow \infty}{\rightarrow} 0.
\end{align*}
\end{theorem}
\begin{proof}
First, based on the convergence theorem in \cite{GEE1}, under the DC-SBM model, the un-normalized embedding for vertex $i$ of class $\mathbf{Y}_i$ satisfies
\begin{align*}
\mathbf{Z}_m(i,k) \stackrel{n \rightarrow \infty}{\rightarrow} \theta_i E(F_\theta) \mathbf{B}_{m}(\mathbf{Y}_i,k).
\end{align*}

It is important to note that the above conditioning assumes knowledge of everything about vertex $i$, including the degree $\theta_i$ and class label $\mathbf{Y}_i$. It follows that the Euclidean norm satisfies
\begin{align*}
\|\mathbf{Z}_m(i,:)\|^2 &= (\sum_{k=1}^{K}\mathbf{Z}_m^2 (i,k)) \\
&\stackrel{n \rightarrow \infty}{\rightarrow} \theta_i^2 E^2(\theta_i) (\sum_{k=1}^{K}\mathbf{B}_{m}^2(\mathbf{Y}_i,k))\\
&= \theta_i^2 E^2(\theta_i) \|\mathbf{B}_{m}(\mathbf{Y}_i,:)\|^2
\end{align*}
Then, the normalized embedding satisfies
\begin{align*}
\mathbf{\tilde{Z}}_m(i,k) = \frac{\mathbf{Z}_m(i,k)}{\|\mathbf{Z}_m(i,:)\|} \stackrel{n \rightarrow \infty}{\rightarrow} \frac{\mathbf{B}_{m}(\mathbf{Y}_i,k)}{\|\mathbf{B}_{m}(\mathbf{Y}_i,:)\|} = \mathbf{\tilde{B}}_{m}(\mathbf{Y}_i,k).
\end{align*}
By concatenating every dimension from $1$ to $K$ within graph $m$, we obtain:
\begin{align*}
\|\mathbf{\tilde{Z}}_m(i,:) - \mathbf{\tilde{B}}_{m}(\mathbf{Y}_i,:)\| \stackrel{n \rightarrow \infty}{\rightarrow} 0.
\end{align*}
The graph fusion embedding is a row-concatenation of $\mathbf{\tilde{Z}}_m(i,:)$ for $m=1,\ldots,M$, so it follows that
\begin{align*}
\|\mathbf{\tilde{Z}}(i, \cdot) - \mathbf{\tilde{B}}(\mathbf{Y}_i,:) \| \stackrel{n \rightarrow \infty}{\rightarrow} 0
\end{align*}
where $\mathbf{\tilde{B}}$ is the row concatenation of every $\mathbf{\tilde{B}}_{m}$.

To verify the rate of convergence, note that
\begin{align*}
\mathbf{W}(i, k) = 1 / n_k
\end{align*} 
when $\mathbf{Y}_i=k$, so
\begin{align*}
\mathbf{Z}_{m}(i, k) & = \mathbf{A}_{m}(i,1:n)\mathbf{W}(1:n,k) = \sum\limits_{j=1,\ldots,n}^{\mathbf{Y}_j=k}\frac{\mathbf{A}_{m}(i,j)}{n_k}.
\end{align*}
Therefore, each dimension is essentially a sample mean using $n_k$ observations per class, with a standard deviation of $O(\frac{1}{\sqrt{n_k}})$. Hence, the convergence rate for dimension $k$ is $O(\frac{1}{\sqrt{n_k}})$, and it is the same rate for any $m$. Now, if we consider the Euclidean norm of all dimensions, i.e., 
\begin{align*}
\|\mathbf{Z}_m(i,:)\| &= \sqrt{(\sum_{k=1}^{K}\mathbf{Z}_m^2 (i,k))},
\end{align*} 
its convergence rate is determined by the slowest across all dimensions, which ends up being $O(\max_{k=1,\ldots,K}\frac{1}{\sqrt{n_k}})$. Note that the probabilistic model assumed that all classes are non-trivial and $\pi_k >0$ for all $k$, it follows that $n_k=O(n)$ for all $k$, and the overall convergence rate can be simplified to $O(\frac{1}{\sqrt{n}})$.
\end{proof}

\begin{theorem}
Suppose that $\{\mathbf{A}_{m}, m=1,\ldots,M\}$ follows the DC-SBM model in Section~\ref{sec1}. As the number of training vertices increases to infinity, the graph fusion embedding achieves asymptotically perfect vertex classification, i.e., 
\begin{align*}
\lim_{n\rightarrow \infty }L_{n}(\mathbf{A}_{1},\ldots,\mathbf{A}_{M}) = 0,
\end{align*}
if and only if there are no repeating rows in $\mathbf{\tilde{B}}$.
\end{theorem}
\begin{proof}
First, it is important to note that the convergence described in Theorem~\ref{thm1} applies to both training vertices and testing vertices, as both use the same $W$ matrix that is based on the labels of the training vertices only. Consequently, the embedding for a testing vertex of class $k$ asymptotically equals the embedding for all training vertices of class $k$, which in turn converges to $\mathbf{\tilde{B}}(k,:)$.

Perfect classification is achievable only when the vertex embedding of each class are completely distinct and well-separated from the vertex embedding of all other classes. Consequently, the testing vertices can be classified perfectly if and only if $\mathbf{\tilde{B}}(k,:) \neq \mathbf{\tilde{B}}(l,:)$ for all pairs of classes $k \neq l$. This condition is met if and only if every row of $\mathbf{\tilde{B}}$ is unique.
\end{proof}

\begin{theorem}
Suppose that $\{\mathbf{A}_{m}, m=1,\ldots,M\}$ follows the DC-SBM model in Section~\ref{sec1}. For any subset of graphs $M_1 \leq M$, the classification error using the fusion embedding satisfies:
\begin{align*}
\lim_{n\rightarrow \infty }L_{n}(\mathbf{A}_{1},\ldots,\mathbf{A}_{M}) - L_{n}(\mathbf{A}_{1},\ldots,\mathbf{A}_{M_1}) \leq 0
\end{align*}
for sufficiently large sample size $n$.
\end{theorem}
\begin{proof}
As the graph fusion embedding converges to the block probability matrix, it suffices to consider the concatenated block matrices. Let $\mathbf{\tilde{B}}^{M_1} \in \mathbb{R}^{K \times M_1 K}$ denote the concatenated block matrix for $M_1$ graphs, and $\mathbf{\tilde{B}}^{M-M_1} \in \mathbb{R}^{K \times (M-M_1) K}$ as the concatenated block matrix for the remaining $M-M_1$ graphs, i.e., 
\begin{align*}
\mathbf{\tilde{B}}^{M} = [\mathbf{\tilde{B}}^{M_1}, \mathbf{\tilde{B}}^{M-M_1}].
\end{align*}
The graph fusion embedding using only $M_1$ graphs asymptotically equals $\mathbf{\tilde{B}}^{M_1}$, and the embedding using $M$ graphs asymptotically equals $\mathbf{\tilde{B}}^{M}$.

If there are no repeating rows in $\mathbf{\tilde{B}}^{M_1}$, then the classification is already perfect using the fusion embedding of $M_1$ graphs. Moreover, all $K$ rows will remain unique in $\mathbf{\tilde{B}}^{M}$, ensuring that the classification remains perfect when using the fusion embedding of all $M$ graphs. Therefore, for sufficiently large $n$, using more graphs in the fusion embedding cannot deteriorate the classification performance.

On the other hand, if there exists a pair $k \neq l$ such that 
\begin{align*}
\mathbf{\tilde{B}}^{M_1}(k,:) = \mathbf{\tilde{B}}^{M_1}(l,:), 
\end{align*}
i.e., using the fusion embedding of $M_1$ graphs, vertices in class $k$ cannot be separated from vertices in class $l$. However, if the remaining $M-M_1$ graphs can separate these two classes, we have
\begin{align*}
\mathbf{\tilde{B}}^{M-M_1}(k,:) \neq \mathbf{\tilde{B}}^{M-M_1}(l,:)
\end{align*}
and
\begin{align*}
\mathbf{\tilde{B}}^{M}(k,:) \neq \mathbf{\tilde{B}}^{M}(l,:).
\end{align*}
Namely, these two classes may become perfectly separable using the fusion embedding of all $M$ graphs. Therefore, for sufficiently large $n$, adding additional graphs never deteriorates the classification performance, but has the potential to improve and successfully classify previously mis-classified vertices. Consequently, 
\begin{align*}
\lim_{n\rightarrow \infty }L_{n}(\mathbf{A}_{1},\ldots,\mathbf{A}_{M_1},\ldots,\mathbf{A}_{M}) - L_{n}(\mathbf{A}_{1},\ldots,\mathbf{A}_{M}) \leq 0.
\end{align*}
\end{proof}

\begin{theorem}
Suppose that $\{\mathbf{A}_{m}, m=1,\ldots,M\}$ follows the general graph model in Section~\ref{sec2}. Let $n$ be the number of vertices with known labels. Then, for any vertex $i$ belonging to class $\mathbf{Y}_i$, its graph fusion embedding satisfies:
\begin{align*}
\|\mathbf{\tilde{Z}}(i, :) - \mu(i) \| = O(\frac{1}{\sqrt{n}}) \stackrel{n \rightarrow \infty}{\rightarrow} 0.
\end{align*}
\end{theorem}
\begin{proof}
For each $\mathbf{Z}_{m}(i, k)$:
\begin{align*}
\mathbf{Z}_{m}(i, k) & = \mathbf{A}_{m}(i,1:n)\mathbf{W}(1:n,k) = \sum\limits_{j=1,\ldots,n}^{\mathbf{Y}_j=k}\frac{\mathbf{A}_{m}(i,j)}{n_k}.
\end{align*}
Given vertex $i$, each $\mathbf{A}_{m}(i,1), \mathbf{A}_{m}(i,2), \ldots, \mathbf{A}_{m}(i,n)$ are independently and identically distributed, and they have finite moments. As $n$ increases to infinity, so is $n_k$, and law of large numbers applies:
\begin{align*}
\mathbf{Z}_{m}(i, k) & \stackrel{n \rightarrow \infty}{\rightarrow} E(\mathbf{A}_{m}(i,j) | i, \mathbf{Y}_j=k).
\end{align*}
Then the Euclidean norm satisfies
\begin{align*}
\|\mathbf{Z}_m(i,:)\| &\stackrel{n \rightarrow \infty}{\rightarrow} \|E(\mathbf{A}_{m}(i,j) | i)\|,
\end{align*}
and
\begin{align*}
\mathbf{\tilde{Z}}_m(i,k) = \frac{\mathbf{Z}_m(i,k)}{\|\mathbf{Z}_m(i,:)\|} \stackrel{n \rightarrow \infty}{\rightarrow} \frac{E(\mathbf{A}_{m}(i,j) | i, \mathbf{Y}_j=k)}{\|E(\mathbf{A}_{m}(i,j) | i)\|}.
\end{align*}

By concatenating all dimensions within the $m$th graph, we obtain:
\begin{align*}
\|\mathbf{\tilde{Z}}_{m}(i,:) - \mu_{m}(i)\| \stackrel{n \rightarrow \infty}{\rightarrow} 0.
\end{align*}
Finally, by concatenating all individual graph embedding into the fusion embedding, we have:
\begin{align*}
\|\mathbf{\tilde{Z}}(i,:) - \mu(i)\| \stackrel{n \rightarrow \infty}{\rightarrow} 0.
\end{align*}
Note that the convergence rate can be proved in the same manner as in the proof of Theorem~\ref{thm1}, and thus is not repeated here.
\end{proof}

\begin{theorem}
Suppose that $\{\mathbf{A}_{m}, m=1,\ldots,M\}$ follows the general graph model in Section~\ref{sec2}. As the number of training vertices increases to infinity, the graph fusion embedding achieves asymptotically perfect vertex classification, i.e., 
\begin{align*}
\lim_{n\rightarrow \infty }L_{n}(\mathbf{A}_{1},\ldots,\mathbf{A}_{M}) = 0
\end{align*}
if and only if 
\begin{align}
\label{eq1}
\sum\limits_{1 \leq k < l \leq K}h_{k}(x)h_{l}(x) = 0
\end{align}
for any $x \in \mathbb{R}^{MK}$.

Additionally, for any subset of graphs $M_1 \leq M$, the classification error using the fusion embedding satisfies:
\begin{align*}
\lim_{n\rightarrow \infty }L_{n}(\mathbf{A}_{1},\ldots,\mathbf{A}_{M_1},\ldots,\mathbf{A}_{M}) - L_{n}(\mathbf{A}_{1},\ldots,\mathbf{A}_{M}) \leq 0
\end{align*}
for sufficiently large sample size $n$.
\end{theorem}
\begin{proof}
As the fusion embedding converges to the concatenated class-conditional means, it suffices to focus on the density family $\{h_{k}(x),k=1,\ldots,K\}$ to analyze the classification performance. When all densities are distinct from each other, it implies that vertex embedding for each class are asymptotically separated from embedding of other classes, leading to perfect classification. In other words, the embedding in class $k$ is perfectly distinguishable from the embedding in class $l$ if and only if $h_{k}(x)h_{l}(x)=0$. This condition extends to Equation~\ref{eq1} by considering all possible pairs.

Regarding the synergistic property, let $\{h_{k}^{M_1}, k=1,\ldots,K\}$ represents the densities using $M_1$ graphs, and $\{h_{k}^{M}, k=1,\ldots,K\}$ represents the densities using all $M$ graphs. Assuming Equation~\ref{eq1} is not equal to zero for embedding using $M_1$ graphs, we define:
\begin{align*}
\tau^{M_1}=\{x \in \mathbb{R}^{M_1 K}, \sum\limits_{1 \leq k < l \leq K}h_{k}^{M_1}(x)h_{l}^{M_1}(x)>0\}
\end{align*}
as the set of mis-classified points. An element $u \in \tau^{M_1}$ has at least one pair of $(k,l)$ such that $h_{k}^{M_1}(u)h_{l}^{M_1}(u)>0$, meaning that $u$ may occur in both class $k$ and class $l$, making these two classes not perfectly separable.

Now, upon adding an additional $M - M_1$ graphs, the element $u$ is appended more entries from the additional graphs, leading to $[u, u'] \in \mathbb{R}^{M K}$. Either $[u, u'] \in \tau^{M}$ and it remains a mis-classified point, or $u'$ is perfectly separable based on the additional $M-M_1$ graphs, leading to $h_{k}^{M}([u, u'])h_{k}^{M}([u, u'])=0$ for the fusion embedding using $M$ graphs. On the other hand, any element not in $\tau^{M_1}$ is already separable in the $\mathbb{R}^{M_{1} K}$, and still separable in the larger, appended $\mathbb{R}^{M K}$ space. Therefore, we have:
\begin{align*}
\tau^{M} \subseteq \tau^{M_1},
\end{align*}
indicating that the probability of mis-classification is either the same or reduced as more graphs are included in the fusion embedding.
\end{proof}

\section{Additional Experiments}
\label{sec:sim2}

\subsection{Heterogeneous Graphs}

\begin{figure}[htbp]
	\centering
	\includegraphics[width=1.0\textwidth,trim={0cm 0.5cm 2cm 0.5cm},clip]{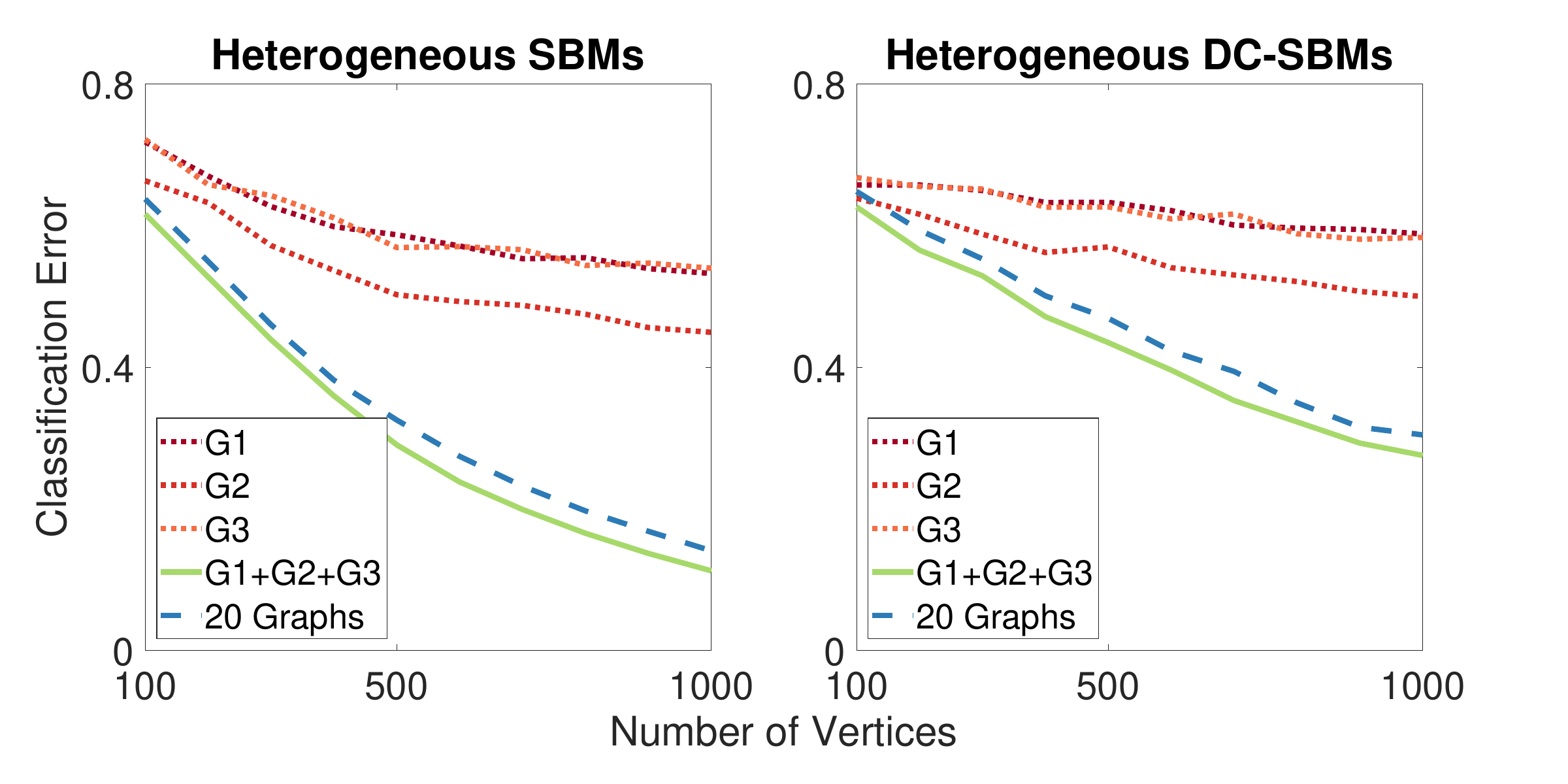}
	\caption{The same experiments as in Figure~\ref{fig1} were conducted using three heterogeneous signal graphs, and the blue dotted line shows the error for adding $17$ noise graphs to the three signal graphs.}
	\label{figS11}
\end{figure}

Figure~\ref{figS11} illustrates the synergistic effect on the same simulations as in Figure~\ref{fig1}, but with heterogeneous graphs. In the SBM setting, we used the same three-graph models as in simulation 1, but we introduced heterogeneity by multiplying each adjacency matrix $A_{m}$ by a random factor $c$, where $c \sim \text{Uniform}(0,0.5)$ for $m=1,2,3$. Additionally, we included $17$ noise graphs that are independent of the labels, with $A_{m}^{h}=A_{m} * c$, where $c \sim \text{Uniform}(1,2)$ for $m>3$. Consequently, the degrees of the vertices are randomized and no longer the same across different graphs, and the noise graphs have larger degrees than the signal graphs. Despite this heterogeneity, Figure~\ref{figS11} clearly shows the same synergistic effect: using all three graphs yields better performance than using any single graph, and the inclusion of additional noise graphs only minimally deteriorates the error. This observation holds true for the DC-SBM setting as well. Note that we omitted the two-graph fusion embedding in the figure for clearer visualization.

\subsection{Alternative Classifiers}

\begin{figure}[htbp]
	\centering
	\includegraphics[width=1.0\textwidth,trim={0cm 0.5cm 2cm 0.5cm},clip]{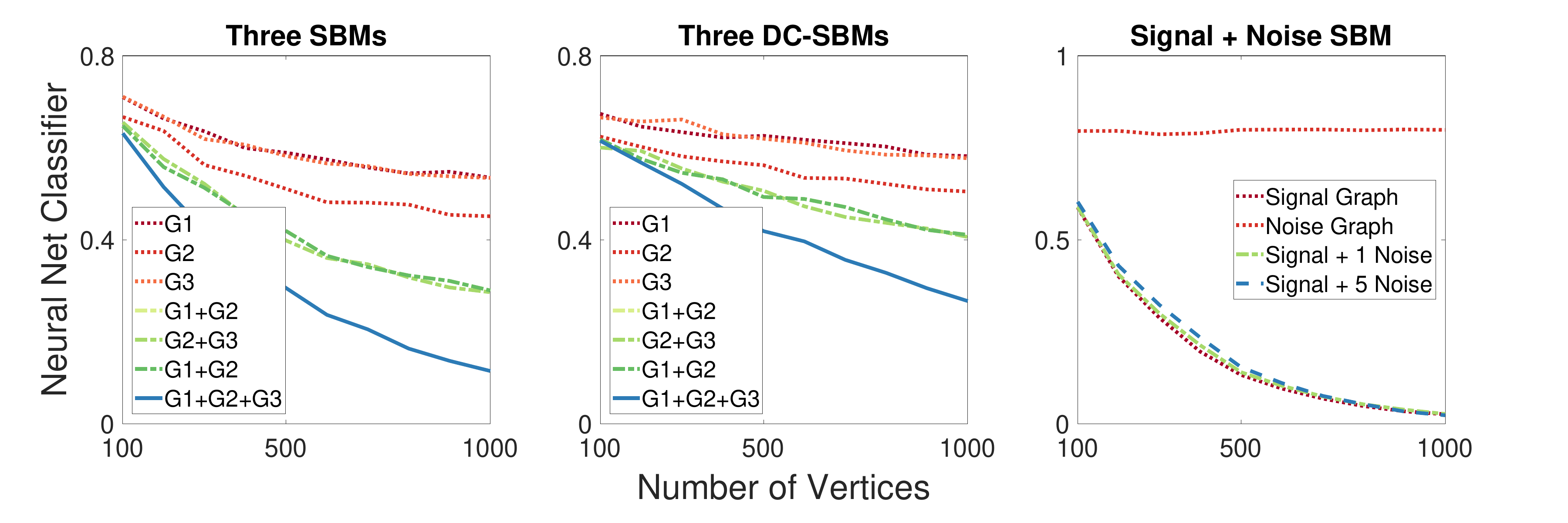}
 \includegraphics[width=1.0\textwidth,trim={0cm 0.5cm 2cm 0.5cm},clip]{FigFusion1Ind2.pdf}
 \includegraphics[width=1.0\textwidth,trim={0cm 0.5cm 2cm 0.5cm},clip]{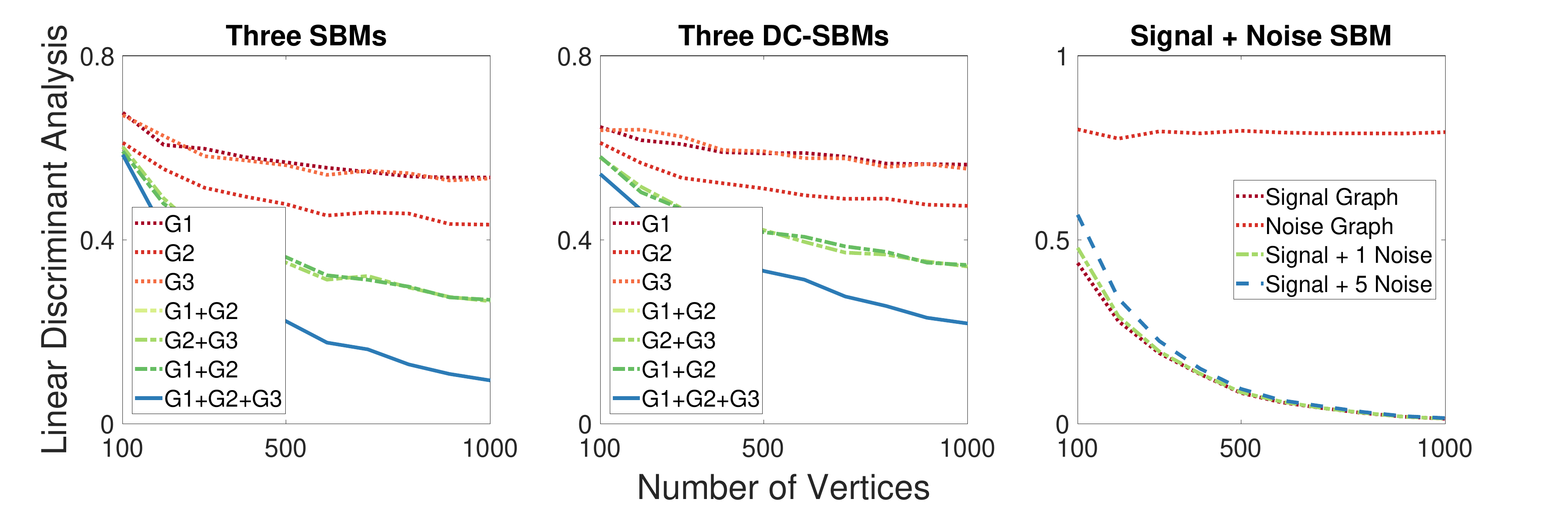}
 \caption{The same experiments as in Figure~\ref{fig1} were conducted using three different classifiers: neural network (first row), nearest-neighbor (second row), and linear discriminant (third row).}
\label{figS12}
\end{figure}

To show that the synergistic effect occurs regardless of the classifier being used, Figure~\ref{figS12} shows the classification results on the fusion embedding using three different classifiers: a two-layer neural network with a neuron size of $100$, the 5-nearest-neighbor classifier (as used in the main paper), and linear discriminant analysis. While there are slight variations in the actual classification errors among the classifiers, the synergistic effect remains consistent across all of them, i.e., including more graphs consistently improves the classification error, and the presence of noise graphs does not deteriorate the error for large sample sizes.

\subsection{Clustering}

In this subsection, we introduce an experimental clustering algorithm for multiple graphs in the absence of ground-truth labels. The approach parallels the clustering algorithm designed for a single graph \cite{GEEClustering}.

Since ground-truth labels are unavailable, we begin by randomly initializing a label vector. This randomized label vector, along with the multiple graphs, serves as input to the main method outlined in Section~\ref{sec:method}. Subsequently, we apply K-means clustering on the output embedding to obtain a new label vector. This new label vector, along with the multiple graphs, is then fed back into the main algorithm iteratively. This iterative process continues until either the new and previous labels are identical, or a pre-defined iteration limit is reached (set to $20$ in our experiments).

The clustering performance is evaluated using the adjusted rand index (ARI), a popular metric for measuring the similarity between two label vectors of the same size. The ARI ranges from $-\infty$ to $1$, where a higher positive value indicates better match quality and a score of $1$ denotes a perfect performance \cite{Rand1971}. Table~\ref{tableA1} presents the clustering performance on individual graphs, all three signal graphs, and all three signal graphs plus $17$ noise graphs, based on the heterogeneous DC-SBM models in Figure~\ref{figS11}.

It is evident from the results that incorporating more graphs leads to improvements in clustering performance. Using all three graphs yields a higher ARI, while the inclusion of noise graphs only minimally affects the performance. Although the synergistic effect appears to extend to the clustering task, there are several areas warranting further investigation. Firstly, due to the iterative nature and random label initialization of the clustering version, a closed-form analysis of embedding convergence is not feasible in the same manner. Secondly, as unsupervised learning poses greater challenges compared to supervised learning, achieving comparable clustering performance requires larger sample sizes. Lastly, as randomization is involved, the algorithm is inherently more volatile and susceptible to noise, as evidenced by the high standard deviation. Therefore, while the clustering performance is promising, further research and algorithm refinement are necessary in the clustering scenario.
\begin{table}[htbp]
\renewcommand{\arraystretch}{1.3}
\centering
{\begin{tabular}{c||c|c|c|c|c}
 \hline
 & G1 & G2 & G3 & G1+2+3 & G1+2+3+Noise  \\
\hline
$n=1000$ & $0.03 \pm 0.02$ & $\textbf{0.11} \pm 0.05$   & $0.02 \pm 0.02$  & $0.04 \pm 0.02$ & $0.04 \pm 0.02$\\
\hline
$n=2000$ & $0.15 \pm 0.03$ & $0.30 \pm 0.04$   & $0.14 \pm 0.03$  & $\textbf{0.40} \pm 0.10$ & $\textbf{0.40} \pm 0.10$\\
\hline
$n=3000$  & $0.21 \pm 0.02$ & $0.36 \pm 0.03$   & $0.21 \pm 0.02$  & $0.67 \pm 0.17$ & $\textbf{0.68} \pm 0.17$\\
\hline
$n=4000$  & $0.23 \pm 0.02$ & $0.39 \pm 0.03$   & $0.23 \pm 0.01$  & $\textbf{0.88} \pm 0.10$ & $0.85 \pm 0.13$\\
\hline
\end{tabular}
\caption{The table displays the adjusted rand index (ARI) for the clustering version of the graph fusion embedding, based on the simulations in the right panel of Figure~\ref{figS11}. We conducted $20$ Monte Carlo replicates and report the average ARI along with the standard deviations.}
\label{tableA1}
}
\end{table}

\subsection{Additional Benchmarks}

In this section, we reproduce the simulations and experiments from the main paper for the Omnibus, MASE, and USE methods. Since these methods rely on spectral embedding, they require selecting a dimension parameter $d$. We evaluate all possible values of $d$ from $1$ to $30$ and report the best error achieved. The classifier used throughout is the 5-nearest-neighbor classifier, consistent with the main paper. 



\begin{figure}[htbp]
	\centering
	\includegraphics[width=1.0\textwidth,trim={0cm 0.5cm 2cm 0.5cm},clip]{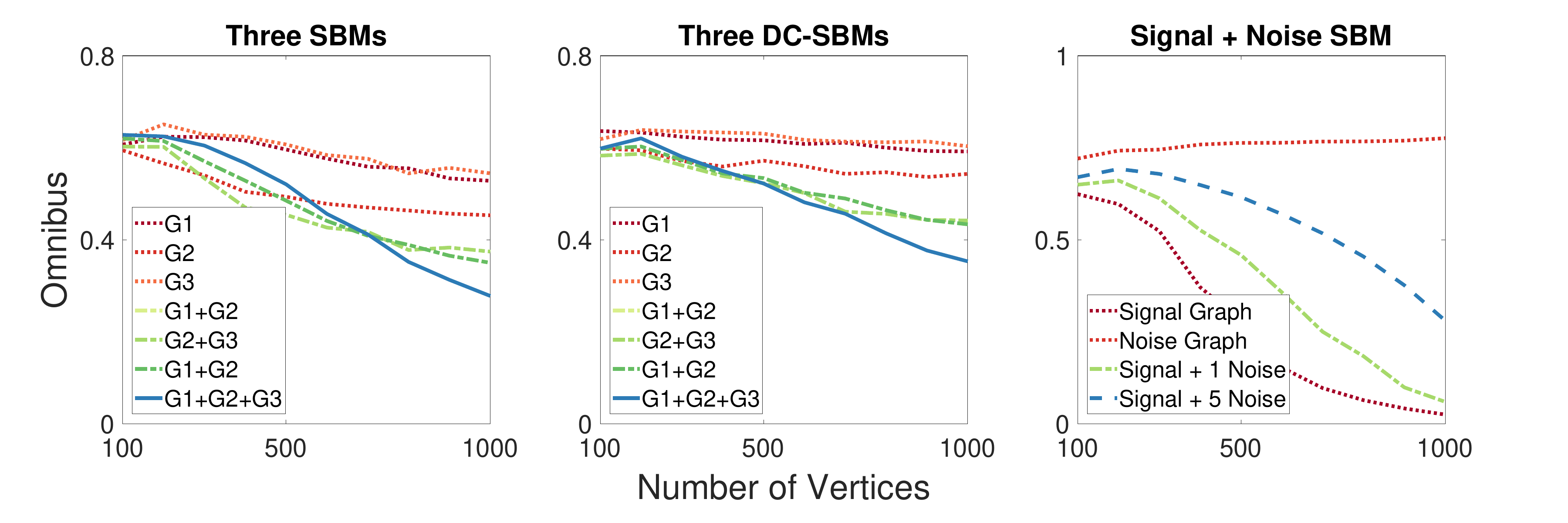}
 \includegraphics[width=1.0\textwidth,trim={0cm 0.5cm 2cm 0.5cm},clip]{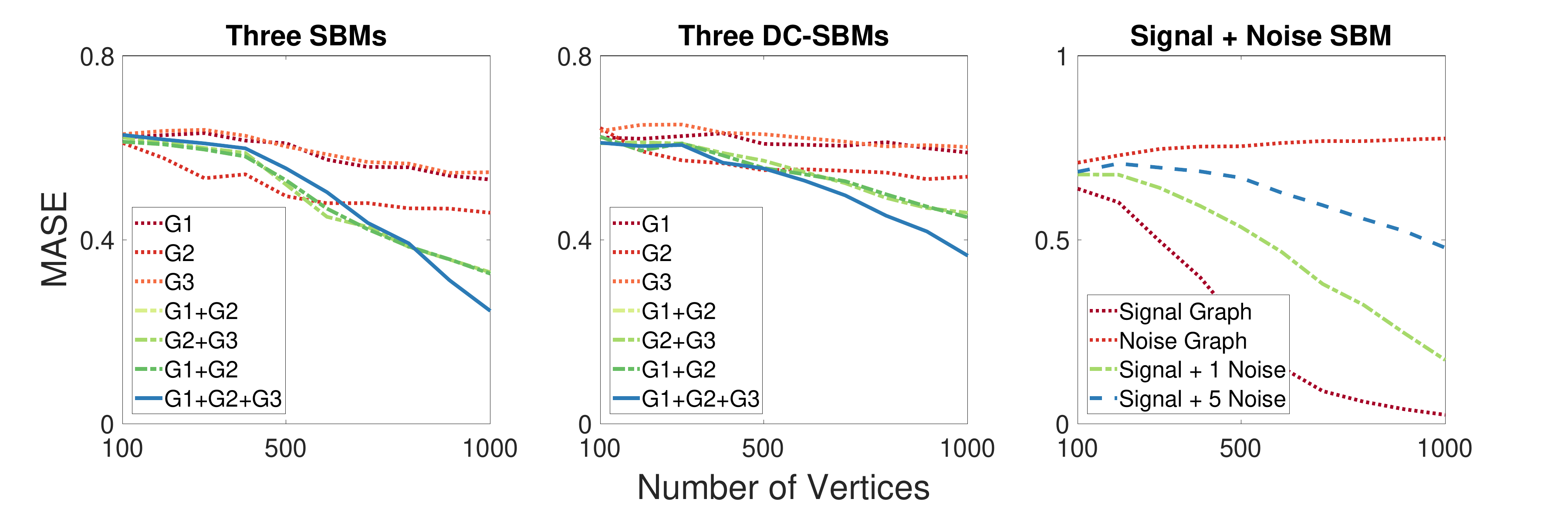}
 \includegraphics[width=1.0\textwidth,trim={0cm 0.5cm 2cm 0.5cm},clip]{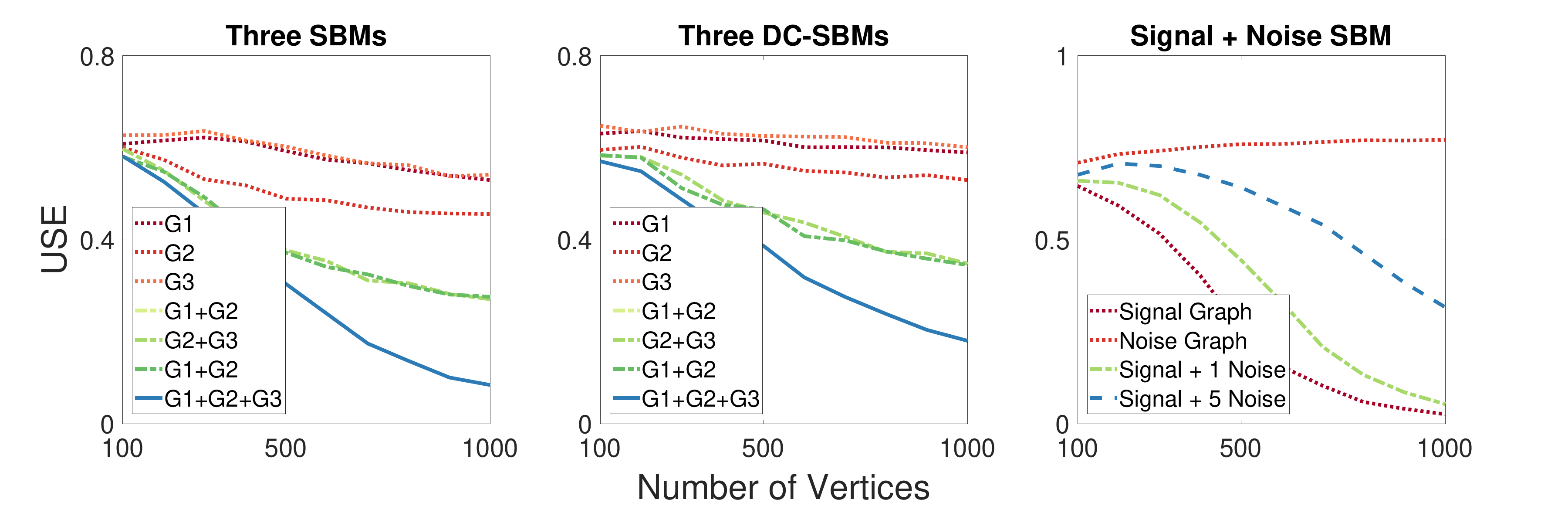}
	\caption{Same experiments as Figure~\ref{fig1} using Omnibus (first row), MASE (second row), and USE (last row).}
	\label{figS1}
\end{figure}

The simulation Figure~\ref{figS1} demonstrate that all three methods are capable of improving classification errors when the additional graph contains signal for the labels. Among the three methods, USE performs the best and has a similar error as the fusion embedding in the main paper. However, in simulation 3, all three methods show a significant degradation in error, especially when five independent graphs are included. 

For the real data experiments presented in the remaining tables, except for the C-elegans data, all three methods exhibit significantly worse actual classification error compared to the fusion embedding and do not demonstrate the synergistic effect.

\begin{table}[htbp]
\renewcommand{\arraystretch}{1.3}
\centering
{\begin{tabular}{c||c|c|c}
 \hline
\textbf{Omnibus} & Graph 1 & Graph 2 & Graph 1+2  \\
\hline
C-elegans & $38.7\% \pm 1.3\%$ & $43.5\% \pm 1.2\%$   & $35.0\% \pm 1.4\%$  \\
\hline
Cora & $31.6\% \pm 0.7\%$ & $41.2\% \pm 0.9\%$     &  $30.7\% \pm 0.5\%$  \\
\hline
Citeseer  & $45.6\% \pm 0.7\%$  & $37.5\% \pm 0.5\%$   &  $33.7\% \pm 0.5\%$  \\
\hline
COIL-RAG  & $76.2\% \pm 0.9\%$ & $36.6\% \pm 0.5\%$    &  $23.9\% \pm 0.4\%$ \\
\hline
IMDB & $19.8\% \pm 0.2\%$     &  $22.2\% \pm 0.2\%$  & $11.8\% \pm 0.2\%$  \\
\hline
\hline
\textbf{MASE} & Graph 1 & Graph 2 & Graph 1+2  \\
\hline
C-elegans & $39.0\% \pm 1.3\%$ & $43.6\% \pm 1.3\%$   & $33.6\% \pm 1.3\%$  \\
\hline
Cora & $31.3\% \pm 0.4\%$ & $41.7\% \pm 0.4\%$     &  $38.2\% \pm 0.6\%$  \\
\hline
Citeseer  & $45.8\% \pm 0.4\%$  & $37.6\% \pm 0.3\%$   &  $36.6\% \pm 0.3\%$  \\
\hline
COIL-RAG  & $77.5\% \pm 0.9\%$ & $36.3\% \pm 0.5\%$    &  $23.7\% \pm 0.4\%$ \\
\hline
IMDB & $19.8\% \pm 0.2\%$     &  $22.2\% \pm 0.2\%$  & $45.6\% \pm 0.2\%$  \\
\hline
\hline
\textbf{USE} & Graph 1 & Graph 2 & Graph 1+2  \\
\hline
C-elegans & $40.1\% \pm 1.5\%$ & $44.1\% \pm 1.5\%$   & $38.2\% \pm 1.5\%$  \\
\hline
Cora & $31.6\% \pm 0.7\%$ & $41.2\% \pm 0.9\%$     &  $30.7\% \pm 0.5\%$  \\
\hline
Citeseer  & $46.4\% \pm 0.5\%$  & $37.3\% \pm 0.3\%$   &  $36.9\% \pm 0.3\%$  \\
\hline
COIL-RAG  & $76.9\% \pm 0.9\%$ & $35.1\% \pm 0.5\%$    &  $35.0\% \pm 0.4\%$ \\
\hline
IMDB & $19.8\% \pm 0.2\%$     &  $22.0\% \pm 0.3\%$  & $12.6\% \pm 0.1\%$  \\
 \hline
\end{tabular}
\caption{Same evaluation as in Table~\ref{table1} using Omnibus, MASE, and USE.}
\label{tableA2}
}
\end{table}

\begin{table*}[htbp]
\renewcommand{\arraystretch}{1.3}
\centering
\scalebox{0.9}{
{\begin{tabular}{c||c|c|c|c}
\hline
\textbf{Omnibus} & English Text & English Hyperlinks & French Text & French Hyperlinks  \\
\hline
One Graph & $21.3\% \pm 0.5\%$  & $22.3\% \pm 0.5\%$ & $41.2\% \pm 0.7\%$ & $45.5\% \pm 0.6\%$ \\
\hline
Two Graphs & $19.4\% \pm 0.4\%$ (1+2)  & $39.3\% \pm 0.6\%$ (3+4) & $41.5\% \pm 0.6\%$ (1+3) & $44.4\% \pm 0.5\%$ (2+4) \\
\hline
Three Graphs & $34.2\% \pm 0.6\%$ (-4)  & $37.5\% \pm 0.6\%$ (-3) & $39.7\% \pm 0.6\%$ (-2) & $39.6\% \pm 0.6\%$ (-1) \\
\hline
All & \multicolumn{4}{c}{$36.5\% \pm 0.5\%$} \\
\hline
\hline
\textbf{MASE} & English Text & English Hyperlinks & French Text & French Hyperlinks  \\
\hline
One Graph & $21.8\% \pm 0.4\%$  & $23.0\% \pm 0.6\%$ & $43.5\% \pm 0.6\%$ & $47.2\% \pm 0.5\%$ \\
\hline
Two Graphs & $19.0\% \pm 0.5\%$ (1+2)  & $40.8\% \pm 0.4\%$ (3+4) & $41.3\% \pm 0.5\%$ (1+3) & $44.2\% \pm 0.7\%$ (2+4) \\
\hline
Three Graphs & $38.1\% \pm 0.7\%$ (-4)  & $39.9\% \pm 0.6\%$ (-3) & $39.3\% \pm 0.7\%$ (-2) & $39.5\% \pm 0.6\%$ (-1) \\
\hline
All & \multicolumn{4}{c}{$37.2\% \pm 0.5\%$} \\
\hline
\hline
\textbf{USE} & English Text & English Hyperlinks & French Text & French Hyperlinks  \\
\hline
One Graph & $21.3\% \pm 0.4\%$  & $22.1\% \pm 0.7\%$ & $41.4\% \pm 0.7\%$ & $45.1\% \pm 0.6\%$ \\
\hline
Two Graphs & $19.2\% \pm 0.4\%$ (1+2)  & $39.9\% \pm 0.6\%$ (3+4) & $39.9\% \pm 0.7\%$ (1+3) & $43.6\% \pm 0.6\%$ (2+4) \\
\hline
Three Graphs & $38.8\% \pm 0.8\%$ (-4)  & $41.3\% \pm 0.7\%$ (-3) & $39.0\% \pm 0.6\%$ (-2) & $39.6\% \pm 0.6\%$ (-1) \\
\hline
All & \multicolumn{4}{c}{$38.7\% \pm 0.6\%$} \\
\hline
\end{tabular}}}
\caption{Same evaluation as in Table~\ref{table3} using Omnibus, MASE, and USE.}
\label{tableA3}
\end{table*}

\end{document}